\documentclass{sig-alternate-10pt}

\usepackage{pslatex}
\usepackage{balance}
\usepackage{xcolor}
\usepackage[margin=1in]{geometry}

\newcommand{\spara}[1]{\smallskip\noindent{\bf #1}}

\usepackage{hyperref}

\begin{document}

\title{LLM+KG@VLDB'24 Workshop Summary}

\author{Arijit Khan$^1$ \,\, Tianxing Wu$^2$ \,\, Xi Chen$^3$  \\
\affaddr{$^1$Aalborg University, Denmark \,\, $^2$Southeast University, China \,\, $^3$Platform and Content Group, Tencent, China} \\
\email{$^1$arijitk@cs.aau.dk \,\, $^2$tianxingwu@seu.edu.cn \,\, $^3$jasonxchen@tencent.com}
}

\maketitle

\vspace{-1in}

\begin{abstract}
The unification of large language models (LLMs) and knowledge graphs (KGs) has emerged as a hot
topic. At the LLM+KG’24 workshop, co-located with VLDB 2024 in Guangzhou, China, the key theme explored was important data management challenges and opportunities due to the effective interaction between
LLMs and KGs. The  report outlines major directions and approaches presented by various
speakers during the 
workshop.
\end{abstract}

\vspace{-2mm}
\section{Introduction}
\label{sec:introduction}

\medskip
\medskip

LLMs, a relatively newer form of generative AI, have become ubiquitous, revolutionizing natural language processing with applications ranging from
solving problems, streamlining workflows, augmenting analytics, code synthesis, to accessing information via conversational functionality, e.g., Copilots  and digital assistants. LLMs are skilled at learning stochastic language patterns as parametric knowledge, and thus predicting next tokens for the given contexts. However, LLMs may lack consistent knowledge representations. Hence, they experience hallucinations and generate unreliable or factually incorrect outputs. KGs can offer external, factual, and up-to-date knowledge to LLMs via, e.g., retrieval augmented methods,  improving the
LLMs' accuracy, consistency, and transparency. On the other hand, LLMs can also facilitate data curation, knowledge extraction, KG creation, completion, embedding, and various downstream tasks over KGs such as recommendation and question answering (QA). Furthermore, the unification of LLMs and KGs creates new data management opportunities and challenges in consistency, scalability, knowledge editing, privacy, fairness,
explainability, data regulations, human-in-the-loop, software-hardware collaboration, cloud-based solutions,
and AI-native databases.

\smallskip

The LLM+KG'24 ambition was to provide a unique platform to researchers and practitioners for presentation of the latest research results, new technology developments and applications, as well as outline the vision for next-generation solutions in the trending topic of unifying LLMs+KGs. The workshop also aims at discussing what interesting opportunities are awaiting for the data management researchers in this greener pasture.

The full-day workshop included 3 keynote talks on the synergies between LLMs and KGs, 1 industrial invited talk on GraphRAG \cite{graphrag}, 9 peer-reviewed research papers from different countries in North and South America, Europe, Asia, and Africa, and a panel discussion on the unification of LLMs, KGs, and Vector databases (Vector DBs). The detailed program is available at \cite{KWC24}.

\vspace{-2mm}
\section{Keynotes}
\label{sec:keynotes}

\medskip
\medskip

The program featured three keynotes by Guilin Qi (Southeast University, China), Haofen Wang (Tongji University, China), and Wei Hu (Nanjing University, China). 

\subsection{Integrating KGs with LLMs: From the Perspective of Knowledge Engineering}
\label{sec:keynote1}

\medskip
\medskip

The first keynote talk on integrating KGs with LLMs from the knowledge engineering point-of-view was given
by Guilin Qi from Southeast University. Prof. Qi started with the enlightening question, {\em ``What is knowledge?''} and shared a number of interesting perspectives. First, according to the Oxford Dictiory, knowledge is the information, understanding, and skills that one gains with education or experience. Second, informally speaking, knowledge can be fact-based,
description of information (e.g., text, image), or skills obtained by practice.
Third, one way to decide whether artificial intelligence (AI) has human intelligence or not could possibly be by the AI's ability to learn and apply knowledge. Fourth, a Knowledge Base (KB) is a collection of knowledge, including documents, images, triples, rules, parameters of neural
networks, etc. Fifth, a KG is a data structure for representing
knowledge using a graph. Prof. Qi further emphasized `KGs as knowledge bases' as follows: ``{\em Knowledge graphs originated from how machines represent knowledge, use graph structures to describe relationships between things, developed in the rise of Web technologies, and landed in applications such as search engine, intelligent QA, and recommender systems''}.


Next, Prof. Qi introduced the fundamentals of language models and
whether they can be used as `parametric knowledge bases' \cite{PetroniRRLBWM19}. He compared the reasoning capabilities of LLMs and KGs, their advantages and disadvantages, and elaborated significant research scopes and practical values due to the complementary nature and mutual enhancements between symbolic knowledge of KGs and parametric knowledge of LLMs.


In the direction of `KGs for LLMs', Prof. Qi discussed how KGs enhance pre-training \cite{Yu0Y022}, fine-tuning \cite{WangZDQZLC24}, inference \cite{WeiH0K23}, prompting \cite{BAS23,KCCYL24}, retrieval/ knowledge augmented generation \cite{XuCGWDWL24}, knowledge editing \cite{ZhengLDFWXC23,YaoWT0LDC023}, knowledge fusion \cite{SXTWLGSG23}, and knowledge validation \cite{HCWQBWP24} of LLMs. In the other direction of `LLMs for KGs', he mentioned knowledge engineering by LLMs, where LLMs can act as both resources (e.g., data augmentation) and enablers (e.g., encoding, reading comprehension, and QA). He also stated several opportunities such as LLMs for entity and relation extraction, triple generation, ontology matching \cite{HertlingP23}, entity alignment \cite{JiangSSXLLGSW24}, knowledge base QA \cite{TanMLLHCQ23}, ontology reasoning \cite{WQLZ24}, and KG reasoning \cite{PanLWCWW24,zhang2024making,wang2024unkr}, among others.


Prof. Qi concluded by underlining interesting opportunities due to LLM+KG integration and the engineering efforts required to work properly, e.g., OpenKG \cite{OpenKG} and new knowledge platforms to support generalizable, trustable, and stable knowledge services. His concluding remark was to look at ``{\em Language as the "form", knowledge as the "heart", and graph as the "skeleton"}.''

\vspace{-1mm}
\subsection{Industry-level KG Platforms for Large-scale, Diverse, and Dynamic Scenarios}
\label{sec:keynote2}

\medskip
\medskip

In the second keynote talk on industry-level knowledge graph platforms, Haofen Wang from Tongji University stated that traditional knowledge semantic frameworks such as RDF/OWL and labeled property graph (LPG) have major limitations in knowledge modeling and management and are often inadequate in modern business scenarios. Prof. Wang provided examples of the Ant Group KG applications in the finance sectors. First, the data sources for KGs have grown tremendously from text to heterogeneous enterprise data, e.g.,  semi-structured/ unstructured user-generated/ professionally generated contents, structured profiles from business operations, transactions, and logs, requiring to implement knowledge hierarchies and
lightweight alignments of diverse sources through programmable methods. Second, knowledge
representations have emerged from binary static structures to multi-dimensional dynamic
associations in temporal and spatial dimensions, therefore deep collaborative information from multiple
aspects of entities, events, concepts, contexts, etc. are required for real-world applications such as merchant management and risk control. In summary, the development of the KG technology does not match the expectations of the new paradigm of an industry-scale, unified, automated knowledge modeling framework for the entire life cycles of businesses, with the ability to evolve and support continuous business iterations.


Next, Prof. Wang introduced the Semantic-enhanced Programmable Graph framework (SPG) developed by the Ant Group and OpenKG \cite{spg} that integrates the structural aspects of LPG with the semantic nature of RDF -- overcoming the semantic complexity of RDF/OWL, while also retaining the simplicity of LPG and
its compatibility with the big data systems. The SPG layered architecture consists of several modules. (i) SPG-Schema is responsible for the schema design. (ii) SPG- Programming, a programmable framework, deals with knowledge construction, knowledge evolution, expert experience projection, and knowledge graph reasoning. (iii)  SPG-Engine is responsible for the execution process of SPG syntax. (iv) SPG-Controller is the control center subsystem, taking care of the control framework, command distribution, and plugin integration. (v) SPG-LLM interacts with LLMs for natural language understanding. 


Prof. Wang concluded by discussing the potentials of SPG and LLM-guided next-generation industry-level cognitive engines, as well as building an AI framework based on the OpenSPG knowledge engine.

\vspace{-1mm}
\subsection{KG-based LLM Fine-tuning} 
\label{sec:keynote3}

\medskip
\medskip

The third keynote talk on KG-enhanced LLM
fine-tuning was given by Wei Hu from
Nanjing University. Prof. Hu emphasized the knowledge gap problem of general-purpose LLMs -- they often lack accurate domain knowledge, resulting in
inaccurate and unreliable outputs, and even difficulty in real-world applications.

Among various knowledge enhancement techniques for LLMs, Prof. Hu focused on an LLM fine-tuning framework with adaptive integration of multi-source KGs, consisting of knowledge extraction, knowledge fusion, and KG-enhanced LLMs. In the field of knowledge extraction, he introduced problems such as domain named
entity recognition, document-level relation extraction \cite{WangWSH22}, continual event extraction \cite{WangW023}, document-level event causality identification \cite{LJZGH23}, and continual relation extraction \cite{WangWH23}. In knowledge fusion, Prof. Hu discussed embedding-based entity alignment \cite{TianS024,SunHXCRH23,GuoZSCHC22,SunHWWQ23}, knowledge transfer \cite{JiaoLWHLBDLHHFY23,WZH24}, adding human-in-the-loop \cite{HuangSCXR023,HuangHBCQ23}, benchmarking, and the OpenEA toolkit \cite{SunZHWCAL20}. In KG-enhanced LLMs fine-tuning, he introduced KnowLA \cite{LuoSZZ024}, a knowledgeable adaptation method for
PEFT (parameter efficient fine-tuning), particularly for LoRA (Low-Rank Adaptation). 
(i) KnowLA with LoRA can align the space of the LLM with the space of KG embeddings, and (ii) KnowLA can activate the parameterized potential knowledge that originally exists in the LLM, even though the used KG does not contain such knowledge.

Prof. Hu concluded with interesting applications of KG-enhanced LLMs in translating configuration files during device replacements in communication networks and unified PEFT+RAG (Retrieval-Augmented Generation).

\section{Industrial Invited Talk}
\label{sec:industry}

\medskip
\medskip

Siwei Gu and Yihang Yu from
NebulaGraph \cite{nebula} delivered an
inspiring industrial talk on GraphRAG \cite{graphrag}, i.e., \textbf{Integrating GenAI with Graph: Innovations and Insights from NebulaGraph}. RAG is a technique to optimize the output of an LLM so that it references an authoritative, up-to-date KB outside of its training corpus before generating a response. Given a user's query, the classic RAG approach uses vector similarity to retrieve semantically similar matches. It also builds offline indexes over embedding vectors to speed-up online retrieval, but partitioning knowledge across chunks can lose global context/ inter-relationships. Connection-oriented retrieval (e.g., join and multi-hop queries) as well as addressing broad, global questions that require synthesizing insights from the entire data can be challenging when the context is spread over multiple chunks.

\smallskip 

To resolve the aforementioned issues,
NebulaGraph launched industry-first GraphRAG \cite{graphrag} -- a technology harnessing the power of knowledge graphs to provide retrieval methods with a more comprehensive contextual understanding and thereby assisting users in obtaining cost-effective, smarter, and more precise search results with an LLM. In particular, it uses a KG to model the external KB, shows the relationships between entities,
which can more accurately understand the query intent, and then uses retrieval enhancement for LLMs. For instance, one can use graph reasoning or subgraph retrieval to find relevant
contexts through relationships.
Users can push domain Knowledge to KG schema and relationships \cite{XuCGWDWL24}.
Furthermore, one can apply graph-based indexing for a more
comprehensive retrieval of context, since graph indexing helps in connecting fragmented knowledge.

Gu and Yu concluded by discussing potential directions about various indexing and retrieval strategies in graphRAG \cite{demo}, node importance finding \cite{GSGYS24}, chain-of-exploration \cite{S24}, and query-focused summarization \cite{ETCBCMTL24}.

\section{Research Papers}
\label{sec:papers}

\medskip
\medskip

The peer-reviewed research papers presented in this workshop can be broadly classified into three categories.

\subsection{LLMs for KGs}
\label{sec:category1}

\medskip
\medskip

KGs are difficult to construct due to the high cost. KG querying is also challenging due to their incompleteness, users requiring to have full knowledge of the query language (e.g., SPARQL, Cypher), and the large and complex KG schema. 
LLMs can assist in KG construction via prompt engineering without huge labeling efforts, and improve the usability and performance of natural language QA with their strong understanding and generalization capabilities. Nie et al. leverage domain-specific knowledge from ontology and Chain-of-Thought prompts to extract higher-quality triples from unstructured text \cite{NHSWZJZS24}. Groves et al. empirically compare in-context learning, fine-tuning, and supervised learning in automated knowledge curation for biomedical ontologies \cite{GWAKHWW24}. Mou et al. explore in-context learning capabilities of GPT-4 for instruction driven adaptive knowledge graph construction, while also proposing a self-reflection mechanism to enable LLMs to critically evaluate their outputs and learn from errors using examples \cite{MLSCD24}. Mustafa et al. use the W3C Open Digital Rights Language (ODRL) ontology and its documentation
to formulate prompts in
large language models and generate usage policies in ORDL from natural language instructions \cite{MNCAQLD24}.

\subsection{KGs for LLMs}
\label{sec:category2}

\medskip
\medskip

LLMs hallucinate due to lack of context or knowledge gap. Offering domain-specific and up-to-date knowledge through KGs can enhance the accuracy, consistency, transparency, and the overall capabilities of LLMs.
Liu et al. propose a collaborative LLMs method for open-set object recognition, incorporating
KGs to alleviate hallucination of LLMs
\cite{LWZCWLW24}. Wang et al.
study a novel infuser-guided knowledge
integration framework to integrate unknown knowledge into LLMs efficiently without unnecessary overlap of known knowledge
\cite{WBWYLCC24}.

\subsection{Unifying LLMs+KGs}
\label{sec:category3}

\medskip
\medskip

The third category of papers simultaneously leverage the factual knowledge of KGs and the parametric knowledge of LLMs to mutually enhance each other. Zhang et al. introduce
OneEdit -- a neural-symbolic prototype system for collaborative
knowledge editing using natural language and facilitating easy-to-use knowledge management with KGs and LLMs \cite{ZXLWTYTYZDS24}.
Khorashadizadeh et al. present a survey on the synergy between LLMs and KGs \cite{KAEIT24}. Cavalleri et al. present the  SPIREX system to extract triples from scientific literature involving RNA molecules \cite{CSPMCRMRCVM24}. They exploit schema constraints in the formulation of LLM prompts and also utilize graph machine learning on an RNA-based KG to assess the plausibility of extracted triples.

\section{Panel}
\label{sec:panel}

\medskip
\medskip

The workshop was concluded with a panel discussion \cite{panel} on the unification of LLMs, KGs, and Vector databases
(Vector DBs). The panelists were Wei Hu (Nanjing University, China), Shreya Shankar (UC Berkeley, USA), Haofen Wang (Tongji University, China), and Jianguo Wang (Purdue University, USA).

\spara{LLMs, KGs, and Vector DBs: Synergy and Opportunities for Data Management.} The LLM+KG’24 chairs first asked some questions.
%
%
{\bf Q1.} What are the synergies among LLMs, Vector DBs, and graph data management including KGs?
%
%
{\bf Q2.} What are the roles of DBs in LLMs + KGs + Vector data management?
%
%
{\bf Q3.} How can LLMs + KGs + Vector data enhance data management?
%
%
{\bf Q4.} What are the significance of human-in-the loop and responsible AI in LLM systems and Vector DBs? How can KGs help in these aspects?
%
%
{\bf Q5.} How can academia + industry partnership and interdisciplinary collaborations advance this field? What would be the roles of benchmarking, open-source models, tools, and datasets?


The panelists added further perspectives to those questions.  While some aspects of these technologies may seem part of the hype cycle, the foundational ideas behind the integration of LLMs, Vector DBs, and KGs are well-grounded in addressing real-world data challenges, and LLMs are definitely a key to genAI. They can reinforce each other by combining structured/ semi-structured and well-curated data for accuracy (e.g., KGs), efficient data retrieval (Vector DBs), and contextual understanding (LLMs), ensuring robust querying, reasoning, and interpretability. Many old DB ideas are relevant around LLMs' self-consistency, thinking step-by-step, etc. \cite{ParameswaranSAJ24}.
For the deploymentment of LLMs in data pipeline, bolt-on data quality constraints for LLM-generated data is crucial \cite{SLAHLZCFPW24}.
LLMs over graph-based applications need both vector- and graph-based RAGs, e.g., consider queries like {\em``What do others say about my papers?''}  or {\em``Find competitors with similar products to mine and analyze their pricing strategies for different products''}. Relational DBs may support efficient vector data management \cite{ZhangLW24}, e.g., PASE is a highly optimized generalized vector
database based on PostgreSQL.

These technologies will enhance databases, knowledge engineering, and data science by enabling more dynamic and responsive search and query responses, facilitating richer interactions with multi-modal data from diverse sources, integrating domain-specific understanding and learning deep semantics. Many potential areas or success stories include NLIDB (natural language interfaces for data bases)/ Text2SQL, query optimization, data curation, neural DBs, self-driving DBs, data education, OpenKG+SPG, and declarative systems for AI workloads (e.g., Palimpzest \cite{LRCCCCFKMV24}, LOTUS \cite{PJGZ24}). The synergy is particularly transformative in domains like personalized healthcare and financial analytics. 

Transparency and explainability are key challenges in this domain. LLMs make mistakes and require guardrails.
Both human-in-the-loop and KGs can align LLMs by providing contextual relevance, factual information, and feedback based on preferences. Ultimately, developing AI systems that adhere to ethical guidelines, emphasizing safety, accountability, fairness, privacy, and transparency is crucial for deploying them in the real world.

This is an interdisciplinary area, and the DB community is well-positioned to own the data pre-processing and validation parts of LLM pipelines \cite{DEEM}. However, encouraging idea exchanges by integrating expertises from fields like DB, ML, NLP, HCI, and CV can drive innovations and create end-to-end solutions/ systems. Fostering academia + industry partnerships would require aligning objectives, e.g., industries can 
offer internships and GPU resources, co-fund initiatives for practical impact and knowledge exchange, while also leading the LLM developments. Benchmarking and providing open source models, tools, and data are important to enhances accessibility, innovation, and community collaboration. Recently, there are also concerns, e.g., many benchmark datasets and empirical studies, 
domain-specific LLMs reporting only “biased” results, etc.

Finally, the panel concluded by discussing open problems such as conducting neural-symbolic reasoning, managing complex, dynamic KGs,
scaling integration and reducing costs,
guardrailing LLMs, ensuring data privacy and compliance, and various engineering challenges.

\section{Future Directions}
\label{sec:conclusion}

\medskip
\medskip

We conclude that there are several
ongoing works in the area of LLMs+KGs, with many open problems, e.g., 

\noindent $\bullet$ \textbf{Integration of Vector and Graph Databases.} Leveraging vector DBs for GraphRAG creates new opportunities such as 
combining graph DBs with vector DBs \cite{liu2025tigervectorsupportingvectorsearch}, using graph DBs as semantic caches of LLMs enabling semantic matching for new graph queries instead of expensive LLM API calls \cite{KhandelwalLJZL20}, optimizing the index creation and similarity search over large-scale graph embeddings, and hardware acceleration.

\noindent $\bullet$ \textbf{Efficient and Explainable GraphRAG.} The efficiency of relevant subgraphs retrieval and raking is challenging in GraphRAG as KGs are large and the context length of LLM is limited. In GraphRAG, KGs can enhance explainability by linking LLM-generated answers to explicit KG relationships, while also acting as guardrails to validate answers against factual knowledge.

\noindent $\bullet$ \textbf{Knowledge Conflict and Dynamic Integration.} Aligning LLMs+KGs is a critical challenge in knowledge engineering since overlap and conflict occur when integrating new knowledge from external sources into LLMs. Incremental updates to KGs and dynamic integration with LLMs are crucial for up-to-date knowledge integration.

The second edition of the workshop LLM+Graph'25 \cite{llmgraph} will be held in conjunction with VLDB 2025 with a broader perspective, since we shall focus on data management for the general topic of LLM+graph computing, rather than only data management for LLM+KG.

\balance

\begin{small}
\bibliographystyle{abbrv}

\begin{thebibliography}{10}

\vspace{6mm}

\bibitem{demo}
{GraphRAG LlamaIndex Workshop}.
\newblock \url{https://colab.research.google.com/drive/1tLjOg2ZQuIClfuWrAC2LdiZHCov8oUbs}.

\bibitem{llmgraph}
{LLM+Graph: The Second International Workshop on Data Management Opportunities in Bringing LLMS with Graph Data}.
\newblock \url{https://seucoin.github.io/workshop/llmg2025/}.

\bibitem{nebula}
{NebulaGraph}.
\newblock \url{https://github.com/vesoft-inc/nebula}.

\bibitem{OpenKG}
{OpenKG}.
\newblock \url{https://github.com/OpenKG-ORG}.

\bibitem{DEEM}
{\em {DEEM '24: Proceedings of the Eighth Workshop on Data Management for End-to-End Machine Learning}}. Association for Computing Machinery, 2024.

\bibitem{spg}
{Ant Group and OpenKG}.
\newblock {Semantic-enhanced Programmable Knowledge Graph (SPG) White paper (v1.0)}.
\newblock \url{https://spg.openkg.cn/en-US}, 2023.

\bibitem{BAS23}
J.~Baek, A.~F. Aji, and A.~Saffari.
\newblock {Knowledge-Augmented Language Model Prompting for Zero-Shot Knowledge Graph Question Answering}.
\newblock In {\em NLRSE@ACL}, 2023.

\bibitem{CSPMCRMRCVM24}
E.~Cavalleri, M.~Soto-Gomez, A.~Pashaeibarough, D.~Malchiodi, H.~Caufield, J.~Reese, C.~J. Mungall, P.~N. Robinson, E.~Casiraghi, G.~Valentini, and M.~Mesiti.
\newblock {SPIREX: Improving LLM-based Relation Extraction from RNA-focused Scientific Literature using Graph Machine Learning}.
\newblock In {\em Workshops at the International Conference on Very Large Data Bases (VLDB)}, 2024.

\bibitem{panel}
X.~Chen, W.~Hu, A.~Khan, S.~Shankar, H.~Wang, J.~Wang, and T.~Wu.
\newblock {Large Language Models, Knowledge Graphs, and Vector Databases: Synergy and Opportunities for Data Management (A Report on the LLM+KG@VLDB24 Workshop's Panel Discussion)}.
\newblock \url{https://wp.sigmod.org/?p=3813}, 2024.

\bibitem{ETCBCMTL24}
D.~Edge, H.~Trinh, N.~Cheng, J.~Bradley, A.~Chao, A.~Mody, S.~Truitt, and J.~Larson.
\newblock {From Local to Global: A Graph RAG Approach to Query-Focused Summarization}.
\newblock {\em CoRR}, abs/2404.16130, 2024.

\bibitem{GWAKHWW24}
E.~Groves, M.~Wang, Y.~Abdulle, H.~Kunz, J.~Hoelscher-Obermaier, R.~Wu, and H.~Wu.
\newblock {Benchmarking and Analyzing In-Context Learning, Fine-tuning and Supervised Learning for Biomedical Knowledge Curation: A Focused Study on Chemical Entities of Biological Interest}.
\newblock In {\em Workshops at the International Conference on Very Large Data Bases (VLDB)}, 2024.

\bibitem{GuoZSCHC22}
L.~Guo, Q.~Zhang, Z.~Sun, M.~Chen, W.~Hu, and H.~Chen.
\newblock {Understanding and Improving Knowledge Graph Embedding for Entity Alignment}.
\newblock In {\em ICML}, pages 8145--8156, 2022.

\bibitem{GSGYS24}
B.~J. Guti{\'{e}}rrez, Y.~Shu, Y.~Gu, M.~Yasunaga, and Y.~Su.
\newblock {HippoRAG: Neurobiologically Inspired Long-Term Memory for Large Language Models}.
\newblock {\em CoRR}, abs/2405.14831, 2024.

\bibitem{HertlingP23}
S.~Hertling and H.~Paulheim.
\newblock {OLaLa: Ontology Matching with Large Language Models}.
\newblock In {\em K-CAP}, pages 131--139, 2023.

\bibitem{HCWQBWP24}
N.~Hu, J.~Chen, Y.~Wu, G.~Qi, S.~Bi, T.~Wu, and J.~Z. Pan.
\newblock {Benchmarking Large Language Models in Complex Question Answering Attribution using Knowledge Graphs}.
\newblock {\em CoRR}, abs/2401.14640, 2024.

\bibitem{HuangHBCQ23}
J.~Huang, W.~Hu, Z.~Bao, Q.~Chen, and Y.~Qu.
\newblock {Deep Entity Matching with Adversarial Active Learning}.
\newblock {\em {VLDB} J.}, 32(1):229--255, 2023.

\bibitem{HuangSCXR023}
J.~Huang, Z.~Sun, Q.~Chen, X.~Xu, W.~Ren, and W.~Hu.
\newblock {Deep Active Alignment of Knowledge Graph Entities and Schemata}.
\newblock {\em Proc. {ACM} Manag. Data}, 1(2):159:1--159:26, 2023.

\bibitem{JiangSSXLLGSW24}
X.~Jiang, Y.~Shen, Z.~Shi, C.~Xu, W.~Li, Z.~Li, J.~Guo, H.~Shen, and Y.~Wang.
\newblock {Unlocking the Power of Large Language Models for Entity Alignment}.
\newblock In {\em ACL}, pages 7566--7583, 2024.

\bibitem{JiaoLWHLBDLHHFY23}
X.~Jiao, W.~Li, X.~Wu, W.~Hu, M.~Li, J.~Bian, S.~Dai, X.~Luo, M.~Hu, Z.~Huang, D.~Feng, J.~Yang, S.~Feng, H.~Xiong, D.~Yu, S.~Li, J.~He, Y.~Ma, and L.~Liu.
\newblock {PGLBox: Multi-GPU Graph Learning Framework for Web-Scale Recommendation}.
\newblock In {\em KDD}, pages 4262--4272, 2023.

\bibitem{KWC24}
A.~Khan, T.~Wu, and X.~Chen.
\newblock {LLM+KG: Data Management Opportunities in Unifying Large Language Models + Knowledge Graphs}.
\newblock In {\em Workshops at the International Conference on Very Large Data Bases {(VLDB)}}, 2024.

\bibitem{KhandelwalLJZL20}
U.~Khandelwal, O.~Levy, D.~Jurafsky, L.~Zettlemoyer, and M.~Lewis.
\newblock {Generalization through Memorization: Nearest Neighbor Language Models}.
\newblock In {\em International Conference on Learning Representations (ICLR)}, 2020.

\bibitem{KAEIT24}
H.~Khorashadizadeh, F.~Z. Amara, M.~Ezzabady, F.~Ieng, S.~Tiwari, N.~Mihindukulasooriya, J.~Groppe, S.~Sahri, and F.~Benamara.
\newblock {Research Trends for the Interplay between Large Language Models and Knowledge Graphs}.
\newblock In {\em Workshops at the International Conference on Very Large Data Bases (VLDB)}, 2024.

\bibitem{KCCYL24}
S.~Ko, H.~Cho, H.~Chae, J.~Yeo, and D.~Lee.
\newblock {Evidence-Focused Fact Summarization for Knowledge-Augmented Zero-Shot Question Answering}.
\newblock {\em CoRR}, abs/2403.02966, 2024.

\bibitem{LRCCCCFKMV24}
C.~Liu, M.~Russo, M.~J. Cafarella, L.~Cao, P.~B. Chen, Z.~Chen, M.~J. Franklin, T.~Kraska, S.~Madden, and G.~Vitagliano.
\newblock {A Declarative System for Optimizing AI Workloads}.
\newblock {\em CoRR}, abs/2405.14696, 2024.

\bibitem{liu2025tigervectorsupportingvectorsearch}
S.~Liu, Z.~Zeng, L.~Chen, A.~Ainihaer, A.~Ramasami, S.~Chen, Y.~Xu, M.~Wu, and J.~Wang.
\newblock {TigerVector: Supporting Vector Search in Graph Databases for Advanced RAGs}.
\newblock {\em CoRR}, 2501.11216, 2025.

\bibitem{LWZCWLW24}
X.~Liu, Y.~Wu, Y.~Zhou, J.~Chen, H.~Wang, Y.~Liu, and S.~Wan.
\newblock {Enhancing Large Language Models with Multimodality and Knowledge Graphs for Hallucination-free Open-set Object Recognition}.
\newblock In {\em Workshops at the International Conference on Very Large Data Bases (VLDB)}, 2024.

\bibitem{LJZGH23}
Y.~Liu, X.~Jiang, W.~Zhao, W.~Ge, and W.~Hu.
\newblock {Dual Graph Convolutional Networks for Document-Level Event Causality Identification}.
\newblock In {\em APWeb-WAIM}, page 114–128, 2023.

\bibitem{LuoSZZ024}
X.~Luo, Z.~Sun, J.~Zhao, Z.~Zhao, and W.~Hu.
\newblock {KnowLA: Enhancing Parameter-efficient Finetuning with Knowledgeable Adaptation}.
\newblock In {\em NAACL}, pages 7153--7166, 2024.

\bibitem{MLSCD24}
Y.~Mou, L.~Liu, S.~Sowe, D.~Collarana, and S.~Decker.
\newblock {Leveraging LLMs Few-shot Learning to Improve Instruction-driven Knowledge Graph Construction}.
\newblock In {\em Workshops at the International Conference on Very Large Data Bases (VLDB)}, 2024.

\bibitem{MNCAQLD24}
D.~M. Mustafa, A.~Nadgeri, D.~Collarana, B.~T. Arnold, C.~Quix, C.~Lange, and S.~Decker.
\newblock {From Instructions to ODRL Usage Policies: An Ontology Guided Approach}.
\newblock In {\em Workshops at the International Conference on Very Large Data Bases (VLDB)}, 2024.

\bibitem{graphrag}
NebulaGraph.
\newblock {NebulaGraph Launches Industry-First Graph RAG: Retrieval-Augmented Generation with LLM Based on Knowledge Graphs}.
\newblock \url{https://www.nebula-graph.io/posts/graph-RAG}, 2023.

\bibitem{NHSWZJZS24}
J.~Nie, X.~Hou, W.~Song, X.~Wang, X.~Zhang, X.~Jin, S.~Zhang, and J.~Shi.
\newblock {Knowledge Graph Efficient Construction: Embedding Chain-of-Thought into LLMs}.
\newblock In {\em Workshops at the International Conference on Very Large Data Bases (VLDB)}, 2024.

\bibitem{PanLWCWW24}
S.~Pan, L.~Luo, Y.~Wang, C.~Chen, J.~Wang, and X.~Wu.
\newblock {Unifying Large Language Models and Knowledge Graphs: A Roadmap}.
\newblock {\em {IEEE} Trans. Knowl. Data Eng.}, 36(7):3580--3599, 2024.

\bibitem{ParameswaranSAJ24}
A.~G. Parameswaran, S.~Shankar, P.~Asawa, N.~Jain, and Y.~Wang.
\newblock {Revisiting Prompt Engineering via Declarative Crowdsourcing}.
\newblock In {\em CIDR}, 2024.

\bibitem{PJGZ24}
L.~Patel, S.~Jha, C.~Guestrin, and M.~Zaharia.
\newblock {LOTUS: Enabling Semantic Queries with LLMs Over Tables of Unstructured and Structured Data}.
\newblock {\em CoRR}, abs/2407.11418, 2024.

\bibitem{PetroniRRLBWM19}
F.~Petroni, T.~Rockt{\"{a}}schel, S.~Riedel, P.~S.~H. Lewis, A.~Bakhtin, Y.~Wu, and A.~H. Miller.
\newblock {Language Models as Knowledge Bases?}
\newblock In {\em EMNLP-IJCNLP}, pages 2463--2473, 2019.

\bibitem{S24}
D.~Sanmartin.
\newblock {KG-RAG: Bridging the Gap Between Knowledge and Creativity}.
\newblock {\em CoRR}, abs/2405.12035, 2024.

\bibitem{SLAHLZCFPW24}
S.~Shankar, H.~Li, P.~Asawa, M.~Hulsebos, Y.~Lin, J.~Zamfirescu-Pereira, H.~Chase, W.~Fu-Hinthorn, A.~G. Parameswaran, and E.~Wu.
\newblock {SPADE: Synthesizing Data Quality Assertions for Large Language Model Pipelines}.
\newblock {\em Proc. VLDB Endow.}, 2024.

\bibitem{SXTWLGSG23}
J.~Sun, C.~Xu, L.~Tang, S.~Wang, C.~Lin, Y.~Gong, H.~Shum, and J.~Guo.
\newblock {Think-on-Graph: Deep and Responsible Reasoning of Large Language Model with Knowledge Graph}.
\newblock {\em CoRR}, abs/2307.07697, 2023.

\bibitem{SunHWWQ23}
Z.~Sun, W.~Hu, C.~Wang, Y.~Wang, and Y.~Qu.
\newblock {Revisiting Embedding-Based Entity Alignment: A Robust and Adaptive Method}.
\newblock {\em {IEEE} Trans. Knowl. Data Eng.}, 35(8):8461--8475, 2023.

\bibitem{SunHXCRH23}
Z.~Sun, J.~Huang, X.~Xu, Q.~Chen, W.~Ren, and W.~Hu.
\newblock {What Makes Entities Similar? {A} Similarity Flooding Perspective for Multi-sourced Knowledge Graph Embeddings}.
\newblock In {\em ICML}, pages 32875--32885, 2023.

\bibitem{SunZHWCAL20}
Z.~Sun, Q.~Zhang, W.~Hu, C.~Wang, M.~Chen, F.~Akrami, and C.~Li.
\newblock {A Benchmarking Study of Embedding-based Entity Alignment for Knowledge Graphs}.
\newblock {\em Proc. {VLDB} Endow.}, 13(11):2326--2340, 2020.

\bibitem{TanMLLHCQ23}
Y.~Tan, D.~Min, Y.~Li, W.~Li, N.~Hu, Y.~Chen, and G.~Qi.
\newblock {Can ChatGPT Replace Traditional KBQA Models? An In-Depth Analysis of the Question Answering Performance of the GPT LLM Family}.
\newblock In {\em ISWC}, pages 348--367, 2023.

\bibitem{TianS024}
X.~Tian, Z.~Sun, and W.~Hu.
\newblock {Generating Explanations to Understand and Repair Embedding-Based Entity Alignment}.
\newblock In {\em ICDE}, pages 2205--2217, 2024.

\bibitem{WBWYLCC24}
F.~Wang, R.~Bao, S.~Wang, W.~Yu, Y.~Liu, W.~Cheng, and H.~Chen.
\newblock {InfuserKI: Enhancing Large Language Models with Knowledge Graphs via Infuser-Guided Knowledge Integration}.
\newblock In {\em Workshops at the International Conference on Very Large Data Bases (VLDB)}, 2024.

\bibitem{wang2024unkr}
J.~Wang, T.~Wu, S.~Chen, Y.~Liu, S.~Zhu, W.~Li, J.~Xu, and G.~Qi.
\newblock {unKR: A Python Library for Uncertain Knowledge Graph Reasoning by Representation Learning}.
\newblock In {\em SIGIR}, pages 2822--2826, 2024.

\bibitem{WQLZ24}
K.~Wang, G.~Qi, J.~Li, and S.~Zhai.
\newblock {Can Large Language Models Understand DL-Lite Ontologies? An Empirical Study}.
\newblock {\em CoRR}, abs/2406.17532, 2024.

\bibitem{WangWH23}
X.~Wang, Z.~Wang, and W.~Hu.
\newblock {Serial Contrastive Knowledge Distillation for Continual Few-shot Relation Extraction}.
\newblock In {\em ACL}, pages 12693--12706, 2023.

\bibitem{WangWSH22}
X.~Wang, Z.~Wang, W.~Sun, and W.~Hu.
\newblock {Enhancing Document-Level Relation Extraction by Entity Knowledge Injection}.
\newblock In {\em ISWC}, pages 39--56, 2022.

\bibitem{WangW023}
Z.~Wang, X.~Wang, and W.~Hu.
\newblock {Continual Event Extraction with Semantic Confusion Rectification}.
\newblock In {\em EMNLP}, pages 11945--11955, 2023.

\bibitem{WangZDQZLC24}
Z.~Wang, Q.~Zhang, K.~Ding, M.~Qin, X.~Zhuang, X.~Li, and H.~Chen.
\newblock {InstructProtein: Aligning Human and Protein Language via Knowledge Instruction}.
\newblock In {\em ACL}, pages 1114--1136, 2024.

\bibitem{WeiH0K23}
Y.~Wei, Q.~Huang, Y.~Zhang, and J.~T. Kwok.
\newblock {KICGPT: Large Language Model with Knowledge in Context for Knowledge Graph Completion}.
\newblock In {\em EMNLP Findings}, pages 8667--8683, 2023.

\bibitem{WZH24}
H.~Wu, X.~Zhu, and W.~Hu.
\newblock {A Blockchain System for Clustered Federated Learning with Peer-to-Peer Knowledge Transfer}.
\newblock {\em Proc. {VLDB} Endow.}, 17(5):966–979, 2024.

\bibitem{XuCGWDWL24}
Z.~Xu, M.~J. Cruz, M.~Guevara, T.~Wang, M.~Deshpande, X.~Wang, and Z.~Li.
\newblock {Retrieval-Augmented Generation with Knowledge Graphs for Customer Service Question Answering}.
\newblock In {\em SIGIR}, pages 2905--2909, 2024.

\bibitem{YaoWT0LDC023}
Y.~Yao, P.~Wang, B.~Tian, S.~Cheng, Z.~Li, S.~Deng, H.~Chen, and N.~Zhang.
\newblock {Editing Large Language Models: Problems, Methods, and Opportunities}.
\newblock In {\em EMNLP}, pages 10222--10240, 2023.

\bibitem{Yu0Y022}
D.~Yu, C.~Zhu, Y.~Yang, and M.~Zeng.
\newblock {JAKET: Joint Pre-training of Knowledge Graph and Language Understanding}.
\newblock In {\em AAAI}, pages 11630--11638, 2022.

\bibitem{ZXLWTYTYZDS24}
N.~Zhang, Z.~Xi, Y.~Luo, P.~Wang, B.~Tian, Y.~Yao, J.~Zhang, S.~Deng, M.~Sun, L.~Liang, Z.~Zhang, X.~Zhu, J.~Zhou, and H.~Chen.
\newblock {OneEdit: A Neural-Symbolic Collaboratively Knowledge Editing System}.
\newblock In {\em Workshops at the International Conference on Very Large Data Bases (VLDB)}, 2024.

\bibitem{zhang2024making}
Y.~Zhang, Z.~Chen, L.~Guo, Y.~Xu, W.~Zhang, and H.~Chen.
\newblock {Making Large Language Models Perform Better in Knowledge Graph Completion}.
\newblock In {\em MM}, pages 233--242, 2024.

\bibitem{ZhangLW24}
Y.~Zhang, S.~Liu, and J.~Wang.
\newblock {Are There Fundamental Limitations in Supporting Vector Data Management in Relational Databases? A Case Study of PostgreSQL}.
\newblock In {\em ICDE}, pages 3640--3653, 2024.

\bibitem{ZhengLDFWXC23}
C.~Zheng, L.~Li, Q.~Dong, Y.~Fan, Z.~Wu, J.~Xu, and B.~Chang.
\newblock {Can We Edit Factual Knowledge by In-Context Learning?}
\newblock In {\em EMNLP}, pages 4862--4876, 2023.

\end{thebibliography}

\end{small}

\end{document}